\documentclass{kluwer}    % Specifies the document style.
\usepackage{graphicx}
\newdisplay{guess}{Conjecture}
\begin{document}
\begin{article}
\begin{opening}         

\title{An evaluation of possible mechanisms for anomalous resistivity in the solar corona}
\author{K.A.P \surname{Singh}\thanks{e-mail: alkendra1978@yahoo.co.in}\thanks{also Indian Institute of Astrophysics, Bangalore- 560034, India}}  
\institute{Department of Applied Physics, Institute of Technology, Banaras Hindu University
Varanasi - 221005, India.}
\author{Prasad \surname{Subramanian}\thanks{e-mail: psubrama@iiap.res.in}}
\institute{Indian Institute of Astrophysics, Bangalore- 560034, India}
 
\runningauthor{Singh \& Subramanian}
\runningtitle{Anomalous Resistivity}

%\date{}

\begin{abstract}
A wide variety of transient events in the solar corona seem to require explanations that invoke fast reconnection. Theoretical models explaining fast reconnection often rely on enhanced resistivity. We start with data derived from observed reconnection rates in solar flares and seek to reconcile them with the chaos-induced resistivity model of Numata \& Yoshida (2002) and with resistivity arising out of the kinetic Alfv\'en wave (KAW) instability. We find that the resistivities arising from either of these mechanisms, when localized over lengthscales of the order of an ion skin depth, are capable of explaining the observationally mandated Lundquist numbers.
\end{abstract}

\end{opening}

\section{Introduction} 
Transient events in the solar atmosphere can often be explained only with models of fast reconnection. While the Petschek (1964) model of reconnection has long been invoked to account for fast reconnection, simulations as well as experiments (e.g., Biskamp 1986; Kulsrud 2001; Ji et al. 1998) seem to indicate that the reconnection geometry might be of the Sweet-Parker kind (Sweet 1958; Parker 1963).
Anomalous resistivity has been often invoked to  
explain fast reconnection (e.g., Biskamp and Welter 1980; Yokoyama and Shibata 1994).
Nakariakov et al. (1999) have invoked an anomalous resistivity that is as much as 6 orders of magnitude larger than the classical value in order to explain the damping of coronal waves (see, however, Klimchuk, Tanner \& DeMoortel 2004 for an alternative explanation). Based on observations of a variety of dynamic events, Dere (1996) concluded that the solar atmosphere is much more resistive than what can be accounted for by classical resistivity. Tsiklauri (2005) has invoked a novel mechanism to account for heating of coronal loops. However, it requires the loops to be comprised of extremely small sub-threads, with dimensions as small as a few proton Larmor radii. If the resistivity is larger than the classical value, it might alleviate this severe requirement on the thickness of the individual strands comprising coronal loops. 

\section{Candidate mechanisms for anomalous resistivity}
Several candidate mechanisms have been proposed for the microscopic origin of anomalous resistivity, such as that due to ion-acoustic turbulence (Bychenkov, Silin \& Uryupin 1988; Uzdensky 2003), due to the kinetic Alfv\'en wave (KAW) instability (Voitekno 1995; Bellan 1999; Bellan 2001), and that arising from chaotic particle motion near the null region (Numata \& Yoshida 2002; 2003). 
In this paper, one mechanism we will examine in detail is that due to Numata \& Yoshida (2002; 2003), where the chaotic motion of particles in the relatively unmagnetized null region mimics collisions and can therefore be used as a basis for deriving an effective anomalous resistivity. The other one we will consider is the current-driven KAW instability.

One of the important bases used for invoking current-driven instabilities such as the ion-acoustic instability or the KAW instability is the presence of 
magnetic field-aligned currents. In turn, the presence of field-aligned currents is inferred from the fact
that solar flares are usually produced from active regions with significant shear, and that there is often a 
significant change in shear following the occurrence of a flare (e.g., Sivaraman, Rausaria \& Aleem 1992).
The premise is that since the reconnecting fields are highly sheared, the classical reconnection geometry with antiparallel
fields annhilating is no longer applicable, and there could well be significant magnetic field components
along the reconnection-induced currents.
 However, such observational inferences arising from two-dimensional pictures of filaments should be treated with caution.
Firstly, as Venkatakrishnan (1993) demonstrates, redeployment of magnetic flux sources relative to the main sunspot(s) and/or emergence of new
flux is a more satisfactory explanation for the observations of Sivaraman, Rausaria \& Aleem (1992). Furthermore, extensive vector magnetogram
observations of flare producing active regions have revealed that apparent photospheric 
magnetic shear is not really
an essential condition for flare production (Wang 1997). Their in-depth study shows that the photospheric magnetic shear does not change after several M-class flares, and it even {\em increases} after the occurrence of large, X-class flares (in fact, it does so for all the X-class flares in their sample). These paradoxes can be
understood only in the context of a three-dimensional reconnection process, of which photospheric shear
provides only a partial, two-dimensional picture.

The crucial difference between a scenario where the resistivity arising out of the current-driven KAW instability would be dominant and one where the chaos-induced resistivity would be so is that the former mechanism can proceed even when the reconnecting magnetic fields are not strictly antiparallel, and magnetic field-aligned currents can therefore be present. 
As discussed above, the three-dimensional geometry of reconnecting fields is not immediately obvious from
current observations. However, although the direct connection to observations of sheared filaments might be simplistic, it is possible that field-aligned currents will exist in reconnection regions. 
It therefore stands to reason that we should consider a general scenario where the reconnecting magnetic
fields need not be exactly antiparallel, where the anomalous resistivity arises out of a current-induced
instability. Of the two current induced instabilities we have mentioned, Bellan (2001) has shown that the KAW instability has a lower threshold than the ion-acoustic one. In addition to the chaos-induced resistivity model of Numata \& Yoshida (2002; 2003), we will therefore also examine the viability of the KAW instability-induced anomalous resistivity using the approach taken
by Voitenko (1995).

\section{Lundquist number comparison}
The Lundquist number gives the ratio of the Lorentz ($J \times B$) force to the force due to resistive magnetic diffusion. We take this to be the figure of merit for evaluating the efficacy of the anomalous resistivity mechanisms we consider. We will derive Lundquist numbers for solar flare events reported in Isobe et al. (2005) and Nagashima \& Yokoyama (2006). We will compare these observationally mandated Lundquist numbers with those derived using the anomalous resistivity mechanisms of Numata \& Yoshida (2002; 2003) and Voitenko (1995).
The macroscopic Lundquist number is defined as
\begin{equation}
S = \frac{V_{A}\,L}{D}\,,
\label{eq1}
\end{equation}
where $V_{A}$ is the Alfv\'en velocity, $L$ is a suitable macroscopic scale length and $D$ is the magnetic diffusivity. In MKS
units, the magnetic diffusivity is defined as
\begin{equation}
D = \frac{\eta}{\mu_{0}}\, \,\,\,\,\, \,\,({\rm m^{2}\,s^{-1}})\,\,,
\label{eq2}
\end{equation}
where $\eta$ is the resistivity and $\mu_{0}$ is the magnetic permeability of free space. Using Eq~(\ref{eq2}) in Eq~(\ref{eq1}) gives
\begin{equation}
S=\frac{V_{A}\,L\,\mu_{0}}{\eta}\,.
\label{eq3}
\end{equation}

For a given transient event in the solar atmosphere, the {\em observed} diffusivity is
\begin{equation}
D_{\rm obs}= \frac{L^{2}}{T} \,\,\,\,\,\,\,\,({\rm m^{2}\, s^{-1}})\,\,,
\label{eq3a}
\end{equation}
where $L$ is the observed lengthscale and $T$ is the observed timescale. This
gives the {\em required} Lundquist number as mandated by the observations,
\begin{equation}
S_{\rm req}= \frac{V_{A}\,L}{D_{\rm obs}}= \frac{V_{A}\,T}{L}\,.
\label{eq3b}
\end{equation}

\subsection{Lundquist numbers from the Numata-Yoshida mechanism}
We now turn our attention to the Lundquist number that can be realised by using the chaos-induced resistivity $\eta_{\rm eff}$ defined in Numata \& Yoshida (2002). 
Using their anomalous resistivity prescription 
\begin{equation}
\eta=\eta_{\rm eff}= \mu_{0}\, \lambda_{i}^{2}\,\omega_{ci}^{2}\,\hat{\nu}_{\rm eff}
\label{eq3c}
\end{equation}
in Eq~(\ref{eq3}) we get
\begin{equation}
S = S_{Y1} = \frac{L\,\omega_{ci}}{V_{A}\,\hat{\nu}_{\rm eff}}= \frac{L}{\hat{\nu}_{\rm eff} \lambda_{i}}\,,
\label{eq4}
\end{equation}
where we have used the following expression for the ion skin depth $\lambda_{i}$:
\begin{equation}
\lambda_{i}= \frac{V_{A}}{\omega_{ci}} \,\,\,\,\,\ \,\,({\rm m})\,\,.
\label{eq5}
\end{equation}
Using the expression $\omega_{ci}=eB/m_{p}$ for the ion cyclotron frequency, we can rewrite Eq~(\ref{eq4}) as 
\begin{equation}
S_{Y1}= \frac{L\,e\,B}{m_{p}\,V_{A}\,\hat{\nu}_{\rm eff}}\,.
\label{eq6}
\end{equation}
The quantity $\hat{\nu}_{\rm eff}$ is the effective collision frequency in units of the ion cyclotron frequency. Numata \& Yoshida (2002) show that, in effect, $\hat{\nu}_{\rm eff}$ is equal to the Alfv\'en Mach number $M_{A}$ of the flow outside the reconnection region. This yields 
\begin{equation}
S_{Y1}= \frac{L\,e\,B}{m_{p}\,V_{A}\,M_{A}}\,.
\label{eq6a}
\end{equation}

The expression for the Lundquist number $S_{Y1}$ given by Eq~(\ref{eq6a}) arises out of using the chaos-induced resistivity $\eta_{\rm eff}$ and assuming that the resistivity is operative over a {\em macroscopic} lengthscale $L$. However, Malyshkin, Linde \& Kulsrud (2005) and Malyshkin \& Kulsrud (2006) suggest that it is not enough for the resistivity to be enhanced for the reconnection to occur at a 
fast rate; the resistivity also needs to be {\em localized} over small lengthscales. If the resistivity is spatially
localized over a lengthscale $\l_{\eta}$, the resulting Lundquist number is obtained by simply using $L=\l_{\eta}$ in Eq~(\ref{eq3}). Numata \& Yoshida's (2002) treatment suggests that the enhanced 
resistivity might be localized over lengthscales comparable to the ion skin
depth $\lambda_{i}$. Using $L=\l_{\eta}= \lambda_{i}$ in (Eq~\ref{eq4}) yields the following expression for the Lundquist number resulting from chaos-induced
resistivity localized over an ion skin depth:
\begin{equation}
S=S_{Y2}= \frac{1}{\hat{\nu}_{\rm eff}} \,.
\label{eq7}
\end{equation}
As mentioned earlier, $\hat{\nu}_{\rm eff}$ can be taken to equal to the Alfv\'en Mach number $M_{A}$ (Numata \& Yoshida 2002), which yields
\begin{equation}
S=S_{Y2}= \frac{1}{\hat{\nu}_{\rm eff}} = \frac{1}{M_{A}}\,.
\label{eq8}
\end{equation}
In writing equation~(\ref{eq8}) it may be noted that we have used the macroscopic Alfv\'en Mach number $M_{A}$, whereas Numata \& Yoshida (2002) have referred to the microscopic Alfv\'en Mach number. 
On the other hand, Lin et al. (2007), have related a microscopic definition of the Lundquist number (the ratio of the resistive diffusion and Alfv\'en timescales, which is equal to the ratio of the width to thickness
of the current sheet), to the macroscopic Alfv\'en Mach number. The use of the macroscopic Alfv\'en number
is primarily because it is the only one that can be observationally estimated.

\subsection{Lundquist numbers from the KAW instability mechanism}
We follow the approach of Voitenko (1995) in evaluating the Lundquist number $S_{KAW}$ arising out of the KAW instability. For conditions applicable to the solar corona (in particular, we note that the Alfv\'en speed they use is similar to the values in table 1), Voitenko (1995) quotes the following approximate value for the magnetic diffusion coefficient $D_{KAW}$ arising from this mechanism:
\begin{equation}
D_{KAW} \simeq 10^{5}\,\,{\rm m^{2}\,s^{-1}} \, .
\label{eq9}
\end{equation}
Following equation~(\ref{eq1}), and since the KAW anomalous resistivity is naturally localized over a thickness of the order of an ion skin depth (Voitenko 1995), we write the Lundquist number $S_{KAW}$ as
\begin{equation}
S_{KAW} = \frac{V_{A}\,\lambda_{i}}{D_{KAW}}\, ,
\label{eq10}
\end{equation}
where we have used $L = \lambda_{i}$.
Equations (\ref{eq9}), (\ref{eq10}), (\ref{eq5}), (\ref{eq3b}) and $\omega_{ci} = eB/m_{p}$ yields
\begin{equation}
\frac{S_{\rm req}}{S_{KAW}} \simeq 10^{5}\,\frac{T}{L} \, \frac{1}{V_{A}}\, \frac{e\,B}{m_{p}}\, .
\label{eq11}
\end{equation}

\section{Results}
We use the formalism developed in the previous section to derive the ratios $S_{\rm req}/S_{Y1}$, $S_{\rm req}/S_{Y2}$ and $S_{\rm req}/S_{KAW}$ for several reconnection events, using observational data given in Nagashima \& Yokoyama (2006). They have compiled a
statistical study of flares observed with the soft X-ray telescope aboard the YOHKOH spacecraft. We have listed the observed lengthscale $L$ and timescale $T$, inferred magnetic field $B$, and reconnection inflow speed $V_{in}$ for each of these events in table 1. The inferred ambient density for each of the events is $n = 10^{15}\,{\rm m^{-3}}$. Using these quantities, we have derived the Alfv\'en speed $V_{A}$ and the Alfv\'en Mach number $M_{A} \equiv V_{in}/V_{A}$.  
We have compared $S_{\rm req}/S_{Y1}$, $S_{\rm req}/S_{Y2}$ and $S_{\rm req}/S_{KAW}$ for each of these events using equations~(\ref{eq3b}), (\ref{eq6a}), (\ref{eq8}) and (\ref{eq11}). $S_{\rm req}$ is the Lundquist number mandated by the observations, while $S_{Y1}$ is the Lundquist number obtained by assuming that the resistivity is due to the Numata-Yoshida mechanism (Numata \& Yoshida 2002; 2003). $S_{Y2}$ is the Lundquist number obtained by assuming that the resistivity is due to the Numata-Yoshida mechanism, and that it is localized over a lengthscale equal to the ion skin depth $\lambda_{i}$. $S_{KAW}$ is the Lundquist number arising from the KAW instability, and is naturally localized over an ion skin depth. 
For all the events, it is evident from table 1 that $S_{Y2}$ is much closer to $S_{\rm req}$ than $S_{Y1}$ is. Unlike Nagashima \& Yokoyama (2006), Dere (1996) does not explicitly list an inflow velocity $V_{in}$ for the events he has considered. Using $V_{in} = L/T$, the Alfv\'en Mach number is $M_{A} = V_{in}/V_{A} = L/V_{A}T$. For each of the events listed in Dere (1996), this yields values of $S_{\rm req}/S_{Y1}$ which are similar to those for the events listed in table 1. However, when this definition of $M_{A}$ is used in the definition of $S_{Y2}$ (equation~\ref{eq8}), it works out to be exactly the same as $S_{\rm req}$ (equation~\ref{eq3b}). 

\section{Summary}
The numbers in table 1 show that the Lundquist number arising from the Numata-Yoshida resistivity localized over an ion skin depth ($S_{Y2}$) as well as that from the KAW instability ($S_{KAW}$) are fairly close to the Lundquist number $S_{\rm req}$ mandated by observations. It is also evident that the resistivity needs not only to be enhanced, but also well localized in order to explain the observations.

\begin{table*}
\begin{tabular}{llllllllll}
\hline
No: & $L$ & $T$ & $B$ & $V_{A}$ & $V_{in}$ & $M_{A}$ & $\frac{S_{\rm req}}{S_{Y1}}$ & $\frac{S_{\rm req}}{S_{Y2}}$ & $\frac{S_{\rm req}}{S_{KAW}}$\\
\hline
1a & $2.56E7$ & $1.32E3$ & $62E-4$ & $4.25E6$ & $4.8E3$ & $1.1E-3$ & $6.73E-8$ &  $0.24$ & $0.72$\\
1b & $2.56E7$ & $1.32E3$ & $116E-4$ & $8.0E6$ & $4.8E3$ & $6.0E-4$ & $6.95E-8$ &  $0.25$ & $0.71$\\
1c & $2.8E7$ & $1.32E3$ & $41E-4$ & $2.8E6$ & $1.3E5$ & $4.7E-2$ & $1.57E-6$ &  $6.204$ & $0.66$\\
2a & $2.94E7$ & $4.8E2$ & $32E-4$ & $2.2E6$ & $1.5E4$ & $7E-3$ & $6.13E-8$ &  $0.25$ & $0.23$\\
2b & $2.94E7$ & $4.8E2$ & $60E-4$ & $4.2E6$ & $1.5E4$ & $3.7E-3$ & $6.3E-8$ & $0.25$ & $0.22$\\
2c & $2.30E7$ & $4.8E2$ & $44E-4$ & $2.1E6$ & $3.2E4$ & $1.5E-2$ & $1.42E-7$ &  $0.66$ & $0.42$\\
3a & $4.12E7$ & $1.2E3$ & $9.0E-4$ & $6.2E5$ & $8.6E3$ & $1.4E-2$ & $4.41E-8$ &  $0.25$ & $0.40$\\
3b & $4.12E7$ & $1.2E3$ & $32E-4$ & $2.3E6$ & $8.6E3$ & $3.9E-3$ & $4.75E-8$ &  $0.26$ & $0.39$\\
3c & $4.0E7$ & $1.2E3$ & $11E-4$ & $9.4E5$ & $6.7E4$ & $7.1E-2$ & $4.46E-7$ & $2.0$ & $0.33$\\
\hline
\end{tabular}
\caption[]{Lundquist number ratios}
% The endnotes section will be placed here.
\theendnotes
The observational data for reconnection events are taken from Nagashima \& Yokoyama (2006). All physical quantities are in MKS units. {\em Column 1}: For each event, we use observational data listed as method 1, method 2, and from Isobe et al. (2005) in Nagashima \& Yokoyama (2006). For instance, 1a refers to method 1, 1b refers to method 2 and 1c refers to data from Isobe et al (2005). {\em Column 2}: The observed lengthscale $L$ of the reconnection event. {\em Column 3}: The observed timescale $T$ of the event. {\em Column 4}: The inferred magnetic field $B$ in the reconnection region. {\em Column 5}: The Alfv\'en speed $V_{A}$. {\em Column 6}: The observed inflow speed $V_{in}$ in the reconnection region. {\em Column 7}: The Alfv\'en Mach number $M_{A}$ ($ = V_{in}/V_{A}$).  {\em Column 8}: Ratio of Lundquist numbers $S_{\rm req}/S_{Y1}$, where $S_{\rm req}$ is given by equation~(\ref{eq3b}) and $S_{Y1}$ by equation~(\ref{eq6a}). {\em Column 9}: Ratio of Lundquist numbers $S_{\rm req}/S_{Y2}$, where $S_{\rm req}$ is given by equation~(\ref{eq3b}) and $S_{Y2}$ by equation~(\ref{eq8}). {\em Column 10}: Ratio of Lundquist numbers $S_{\rm req}/S_{KAW}$ given by equation~(\ref{eq11}).
\end{table*}
\acknowledgements
KAPS acknowledges the support from the University Grants Commission, New Delhi for the award of the senior research fellowship. We acknowledge constructive criticism from an anonymous referee which has helped us significantly improve the paper.

\end{article}
\end{document}